 \documentclass[tablecaption=bottom,pmlr]{jmlr}

\usepackage{booktabs}
\usepackage[load-configurations=version-1]{siunitx} 
\usepackage{tikz}
\usetikzlibrary{quantikz}
\newcommand\inlineeqno{\stepcounter{equation}\ (\theequation)}
\usepackage{bbm}
\usepackage{physics}
\usepackage{dirtytalk}
\usepackage{float}
\usepackage{todonotes}
\usepackage{caption}

\theorembodyfont{\upshape}
\theoremheaderfont{\scshape}
\theorempostheader{:}
\theoremsep{\newline}

\jmlrvolume{148}
\jmlryear{2021}
\jmlrsubmitted{submission date}
\jmlrpublished{publication date}
\jmlrworkshop{NeurIPS 2020 Preregistration Workshop} 

\title{Playing Atari with Hybrid Quantum-Classical Reinforcement Learning}

  \author{\Name{Owen Lockwood} \Email{lockwo@rpi.edu}
   \\ \addr Department of Computer Science \\ Rensselaer Polytechnic Institute, Troy, NY, USA 
   \AND
   \Name{Mei Si} \Email{sim@rpi.edu}
   \\ \addr Department of Cognitive Science \\ Rensselaer Polytechnic Institute, Troy, NY, USA}
   
\begin{document}

\maketitle

\begin{abstract}
Despite the successes of recent works in quantum reinforcement learning, there are still severe limitations on its applications due to the challenge of encoding large observation spaces into quantum systems. To address this challenge, we propose using a neural network as a data encoder, with the Atari games as our testbed. Specifically, the neural network converts the pixel input from the games to quantum data for a Quantum Variational Circuit (QVC); this hybrid model is then used as a function approximator in the Double Deep Q Networks algorithm. We explore a number of variations of this algorithm and     find that our proposed hybrid models do not achieve meaningful results on two Atari games -- Breakout and Pong. We suspect this is due to the significantly reduced sizes of the hybrid quantum-classical systems. 
\end{abstract}


\section{Introduction}

Reinforcement Learning (RL) has advanced dramatically in the last decade. Deep RL has achieved remarkable results on a variety of very complex tasks such as Chess, Go \citep{silver2018general}, StarCraft II \citep{vinyals2019grandmaster}, autonomous navigation \citep{bellemare2020autonomous}, and computer chip design \citep{mirhoseini2021graph}. Games and control tasks are common RL benchmarks due to their structured observations and rewards. Newer RL algorithms are constantly being created to achieve superior performance and to achieve this performance with less training \citep{mousavi2016deep}. 

Quantum computing has also made significant advancements in recent decades. Early work in quantum computing was catalyzed by Shor's algorithm, a polynomial time algorithm for integer factorization with significant cryptographic implications \citep{shor1999polynomial}. However, it has only been in recent years that quantum computing became a realizable technology, with a number of quantum computers being developed that claim to possess an advantage over their classical counterpart \citep{arute2019quantum, zhong2020quantum, arrazola2021quantum}. By leveraging quantum phenomena such as superposition and entanglement, quantum computers can offer computational advantages. 

Quantum machine learning (QML) has attracted an increasing amount of attention in recent years. There is significant potential for theoretical quantum speedups on machine learning tasks, e.g. quantum perceptrons and quantum RL have the potential for $O(\sqrt{N})$ speedups \citep{biamonte2017quantum}. Already work has been done to develop quantum generative adversarial networks \citep{dallaire2018quantum, lloyd2018quantum}, quantum Hopfield networks \citep{rebentrost2018quantum} and quantum support vector machines \citep{rebentrost2014quantum}. Recently, the quantum RL field \citep{dong2008quantum} has been expanding with a variety of approaches. Some recent works utilize Grover iterations \citep{hu2018empirical} and quantum Boltzmann machines \citep{jerbi2019framework} to master simple environments such as Gridworld and CartPole. 

In this work, we want to investigate the potential that quantum computing has to aid with reinforcement learning. We expand upon our previous work \citet{lockwood2020}, which was in turn inspired by \citet{chen2020variational} to use Quantum Variational Circuits (QVC) -- quantum circuits with gates parameterized by learnable values -- in reinforcement learning. In \citep{chen2020variational}, QVCs were used with Double DQN for a deterministic 4x4 Gridworld. \citet{chen2020variational} reported that the parameter space complexity scales linearly with the input space in QVCs, i.e. $O(N)$, which is a significant improvement over the traditional neural network DQN which has parameter space complexity $O(N^2)$. They used computational basis encoding, which involves converting the input into a binary number and flipping a sequence of qubits to represent that binary number. However, this technique is unsuitable when the size of inputs is large or when floating point inputs are involved. \citet{lockwood2020} demonstrated the applicability of using QVCs to environments with larger input spaces by utilizing more efficient encoding schemes. Specifically, the encoding scheme transformed each value in the input into rotations for a single qubit. This means that the number of qubits required is equal to the length of the input. This is feasible when the input is of size 4 (like in CartPole), but not possible for larger input spaces. For Atari, this would require 7,056 qubits for a single frame. This is infeasible, meaning that traditional benchmarks (like Atari) remain inaccessible. Although algorithms exist for optimal (amplitude) encoding schemes, i.e. encoding $2^N$ numbers in N qubits, they require an exponential number of gates (in relation to the input size) which is not only intractable but negates the exponential gains from the amplitude encoding \citep{shende2006synthesis, mottonen2004transformation}.

Previously techniques relied on static encoders, but in this work, we present a solution for the quantum reinforcement learning encoding problem via a learned algorithm. Specifically, we employ a neural network to encoding classical data into quantum circuits. The neural network takes the environment observations as input and outputs operations to encode the information into the quantum circuit. The Atari environments were previously inaccessible due to the dimensionality and size of the required inputs, but are made available by using a neural network encoder as we proposed. 

For evaluating the feasibility of this encoder, and the potential for QML to assist with RL tasks, we propose an empirical study to be conducted on the Noisy Scale Intermediate Quantum (NISQ) \citep{preskill2018quantum} QML simulator TensorFlow-Quantum \citep{broughton2020tensorflow}. In this study, we compare hybrid quantum-classical approaches with purely classical approaches. We apply our techniques to two pixel based Atari OpenAI Gym environments, Breakout and Pong \citep{brockman2016openai}. The input into our proposed models is more than 7,000 times larger than CartPole. The input size of CartPole is 4 and the input size for Atari games is $84 * 84 * 4 = 28224$. CartPole was used in \citep{lockwood2020}, which is considered one of the most complex previous environments. We hope that using a neural network as an encoder will solve the previous problems of encoding and enable us to unlock quantum advantages even for large, high dimensional input spaces. 

\section{Background}

\subsection{Reinforcement Learning}

Reinforcement learning is a form of learning in which at least one agent interacts with an environment with the goal of maximizing a numerical reward signal \citep{sutton2018reinforcement}. A common formalization of RL are Markov Decision Processes (MDPs). The MDP tuple, $\langle \mathcal{S}, \mathcal{A}, P, R, \gamma\rangle$, consists of a set of states $\mathcal{S}$, actions $\mathcal{A}$, the probability of transition from one state to the next $P = P[s_{t+1} = s^{\prime} | s_{t} = s, a_t = a]$, the reward, $R$, and the future reward discount $\gamma$. The goal is to design an agent that with policy $\pi$, $\pi(s_t) = a_t$, such that it maximizes the expected reward, $\mathbbm E \left [\sum_{t=0}^T \gamma^t R(s_t, a_t)|\pi \right ]$. 

Deep Q Networks (DQN) are an off policy and model free algorithm that uses a function approximator to estimate the Q function \citep{mnih2013playing}. The Q function approximator, parameterized by $\theta$, is defined as the expected future reward $Q_\theta(s,a) = \mathbbm{E}_\theta [R | s_0 = s, a_0 = a]$. This can also be defined recursively for easier updates: $Q(s_t,a_t)=r_t+max_{a_{t+1}}Q(s_{t+1},a_{t+1})$, where $Q(s_t,a_t)$ is the Q value of a certain action in a given state at time t. The original Q learning policy is defined for discretized action spaces and is defined as $\pi(s;\theta) = max_aQ(s;\theta)$, i.e. the policy is to choose the action with the largest Q value as approximated by the parameters $\theta$ \citep{watkins1992q}. Updates to this policy are made via the mean squared Bellman error, $L_t(\theta)=\mathbbm E[(r_t+\gamma * max Q(s',a';\theta)-Q(s,a;\theta))^2]$ from which gradients can be calculated \citep{mnih2013playing}. In Q learning with function approximation, failure to converge is a common problem. One source of this problem is the max operation, which leads to over-estimations of the Q value \citep{thrun1993issues}. One solution to this is Double Q learning, used in this work, in which a separate target network is used exclusively for predicting the future Q value inside the max operation \citep{van2015deep}. In this work we use neural networks and QVCs as function approximators that estimate the Q values.

\subsection{Quantum Machine Learning}

Quantum machine learning (QML) is at the intersection of machine learning and quantum computing. It seeks to use quantum computing to obtain quantum advantage on machine learning tasks. Quantum advantages often stem from the abilities of quantum computers to represent and operate on information that scales exponentially with the number of qubits. 

Two of the most important features of quantum mechanics that quantum computing exploits are superposition and entanglement. Unlike in classical computers, where bits are limited to be 0 or 1, quantum bits (qubits) are capable of representing both 0 and 1 simultaneously. This is because the qubit is a quantum mechanical wavefunction $\Psi$ that can be a linear combination of terms, e.g. $\Psi = \alpha |0 \rangle + \beta |1\rangle$. This enables information represented to scale $O(2^N)$ for N qubits, giving an exponential advantage over linearly scaling classical bits. However, it is important to note that only a single value can be obtained from the wavefunction as it 'collapses' once a Hermitian operator (i.e. a measurement) is applied.

Entanglement is a more complex phenomenon resulting from the inseparability of combined wavefunctions. When two qubits are separate (i.e. not entangled) their wavefunction can be mathematically divided into individual wavefunctions. Consider one qubit in a superposition and another in the state $|0\rangle$, the two qubit wavefunction would be: $\Psi = \alpha |00\rangle + \beta |10\rangle$. This can easily be separated into $\Psi = (\alpha |0\rangle + \beta |1\rangle)(|0\rangle)$. However if the two qubits are entangled this is not possible. Consider the two qubit wavefunction, called the Bell State or EPR state, $\Psi = \frac{1}{\sqrt{2}} |00\rangle + \frac{1}{\sqrt{2}} |11\rangle$. If we were to attempt to separate this wavefunction and write it as two distinct wavefunctions that are simply multiplied together, we would see get $\Psi = (a |0\rangle + b |1\rangle)(c |0\rangle + d |1\rangle)$. However, this would require $a*c = \frac{1}{\sqrt{2}}$, $b*d = \frac{1}{\sqrt{2}}$, $a*d = 0$ and $b*c = 0$ which is clearly not possible. This is what is meant by "inseparable". This has important implications because it means that a single operation on one qubit influences all qubits because their wavefunctions are mathematically inseparable. 

\subsubsection{Quantum Variational Circuits}

Special operations, called quantum gates, are required in order to manipulate qubits. There are a number of quantum gates, but the ones relevant to this work are the Pauli rotation gates and the controlled NOT (CNOT) gate. The Pauli rotations gates, $R_x(\theta), R_y(\theta), R_z(\theta)$, rotate around the specified axis $\theta$ radians. Mathematically: $R_\alpha(\theta) = e^{-i\frac{\theta}{2}\sigma_\alpha}$, where $\alpha = X, Y, Z$. The controlled NOT (CNOT) gate is a two qubit gate that can induce entanglement in qubits. The CNOT is not parameterized and is used for entanglement purposes only. The aforementioned $\theta$ are the learnable parameters that are updated via gradient descent. 

Quantum Variational Circuits (QVCs) are a collection of qubits and the set of gates that operate on them \citep{mcclean2016theory}. There are three main components of a QVC: an encoding circuit, a parameterized circuit, and a readout circuit \citep{benedetti2019parameterized}. The encoder converts classical data into quantum data (i.e. parameter free quantum circuits). The parameterized circuit operates on the on the quantum data to produce an approximation of the desired state. Finally, a readout measurement is taken, usually one of the Pauli operators ($X, Y, Z$) is used to extract information from the circuit. In this work we use the Z operator, or the 'computational basis state'. External to the quantum circuit (on a classical computer) a loss function and associated gradients are calculated then the parameters are updated. 

The gradients for QVCs cannot be calculated using the same differentiation techniques as traditional neural networks. Quantum gradients on hardware can be calculated using the parameter shift differentiator. This differentiator is implemented as part of TensorFlow-Quantum package. The rotation that a gate enacts on a qubits can be represented in the form: $U_{i}^{\ell}(\theta_i^{\ell}) = e^{-iaG\theta_i^{\ell}} \inlineeqno$, where $\ell$ is the layer index, $a$ is a constant and $G$ a linear combination of Pauli gates, called a generator \citep{crooks2019gradients}. A QVC is a function of $\theta$, and is equivalent to the expectation value of the readout operator ($\hat{Z}$ in this work). This is written as: $f(\theta) = \langle \Psi_0 | U^{\dag}(\theta) \hat{Z} U(\theta) | \Psi_0 \rangle$, where $\Psi_0$ represents the initial wavefunction \citep{broughton2020tensorflow}. The parameter shift rule states that $\pdv{}{\theta}f(\theta) = \langle \Psi_0 | (\pdv{}{\theta}U^{\dag}(\theta)) \hat{Z} U(\theta) | \Psi_0 \rangle + \langle \Psi_0 | U^{\dag}(\theta) \hat{Z} (\pdv{}{\theta}U(\theta)) | \Psi_0 \rangle \inlineeqno$ \citep{schuld2019evaluating}. Equations $(1)$ and $(2)$ can be combined to yield a differentiation rule: $\pdv{}{\theta}f(\theta) = r[f(\theta+\frac{\pi}{4r}) - f(\theta-\frac{\pi}{4r})] \inlineeqno$ \citep{crooks2019gradients}. In this formula $r$ is a value that can vary between implementations but is often set in relation to the eigenvalues of the operations $e_0, e_1$ where $r = \frac{a}{2}(e_1-e_0)$ . Thus in the case of Pauli gates, $r=\frac{1}{2}$ because the eigenvalues of all Pauli matrices are $-1, 1$. Equation $(3)$ is the parameter shift technique for how to differentiate through a quantum circuit enabling both gradients for the circuit and backpropagation through the circuit. 

The idea behind this work is to circumvent the traditionally hard problem of encoding classical states into quantum circuits with a neural network. Because of the differentiability of the quantum circuit, gradients can be carried through and applied to the neural network encoder. The neural network will output the rotations of gates that transform pure states. This is a significant departure from previous approaches to quantum RL and should enable significantly larger state encoding.  

\section{Approach}

\subsection{Methodology}\label{method}

The algorithm used in this work is Double Deep Q Learning (DDQN) \citep{van2015deep}. As in other works, the only algorithmic differences are the function approximators, the fundamentals of the algorithm remain the same \citep{lockwood2020, chen2020variational}. The simplicity of just replacing the neural network with a QVC or hybrid model has been shown to work in simple applications like CartPole \citep{lockwood2020} and Gridworld \citep{chen2020variational} environments. The setup of the Atari benchmark also remains unmodified, in that the goal is to maximize the reward achieved and the input is the 4 framestacked 84X84 images which has been cropped and grey scaled. 

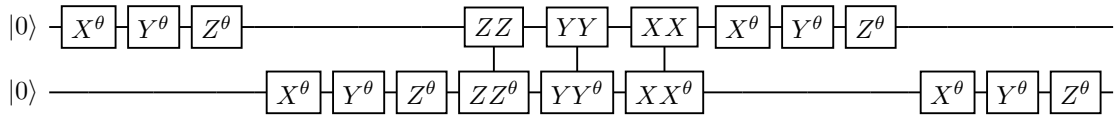
\begin{figure}[h]
    \centering
    \scalebox{0.9}{
    \begin{quantikz}[row sep=0.35cm, column sep=0.185cm]
    \lstick{$\ket{0}$} & \gate{X^{\theta}} & \gate{Y^{\theta}} & \gate{Z^{\theta}} & \qw & \qw & \qw & \qw & \gate{ZZ} & \gate{YY} & \gate{XX} & \gate{X^{\theta}} & \gate{Y^{\theta}} & \gate{Z^{\theta}} & \qw & \qw & \qw & \qw & \qw \\
    \lstick{$\ket{0}$} & \qw & \qw & \qw & \qw & \gate{X^{\theta}} & \gate{Y^{\theta}} & \gate{Z^{\theta}} & \gate{ZZ^\theta}\vqw{-1} & \gate{YY^\theta}\vqw{-1} & \gate{XX^\theta}\vqw{-1} & \qw & \qw & \qw & \qw & \gate{X^{\theta}} & \gate{Y^{\theta}} & \gate{Z^{\theta}} & \qw \\
    \end{quantikz}
    }
    \caption{Two Qubit Unitary}
    \label{fig:conv}
\end{figure}

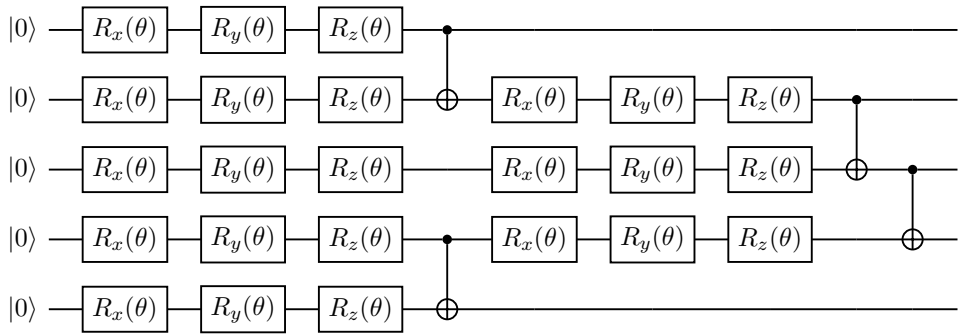
\begin{figure}[h]
    \centering
    \scalebox{0.9}{
    \begin{quantikz}[row sep=0.35cm]
    \lstick{$\ket{0}$} & \gate{R_x(\theta)} & \gate{R_y(\theta)} & \gate{R_z(\theta)} & \ctrl{1} & \qw & \qw & \qw & \qw & \qw & \qw  \\
    \lstick{$\ket{0}$} & \gate{R_x(\theta)} & \gate{R_y(\theta)} & \gate{R_z(\theta)} & \targ{} & \gate{R_x(\theta)} & \gate{R_y(\theta)} & \gate{R_z(\theta)} & \ctrl{1} & \qw & \qw \\
    \lstick{$\ket{0}$} & \gate{R_x(\theta)} & \gate{R_y(\theta)} & \gate{R_z(\theta)} & \qw & \gate{R_x(\theta)} & \gate{R_y(\theta)} & \gate{R_z(\theta)} & \targ{} & \ctrl{1} & \qw \\
     \lstick{$\ket{0}$} & \gate{R_x(\theta)} & \gate{R_y(\theta)} & \gate{R_z(\theta)} & \ctrl{1} & \gate{R_x(\theta)} & \gate{R_y(\theta)} & \gate{R_z(\theta)} & \qw & \targ{} & \qw \\
    \lstick{$\ket{0}$} & \gate{R_x(\theta)} & \gate{R_y(\theta)} & \gate{R_z(\theta)} & \targ{} & \qw & \qw & \qw & \qw & \qw & \qw\\
    \end{quantikz}
    }
    \caption{Single Layer of Parameterized Circuit}
    \label{fig:cir1}
\end{figure}

For the quantum architecture, we use the quantum convolution operation (QCNN) which serves as a quantum parallel to the classical CNN, with the same goal of feature extraction. The QCNN is a parameterized two qubit unitary, i.e. arbitrary entangled rotation, on every set of adjacent qubits \citep{cong2019quantum}. A single two bit unitary operation is shown in Figure \ref{fig:conv}. After the quantum convolutional layers, there are 3 layers of the circuit with the same architecture shown in Figure \ref{fig:cir1}. More qubits can be added to this circuit by expanding either the inner set or outer set and using a CNOT gate to induce entanglement with the rest of the circuit. Note that $R_\alpha(\theta)$ rotates about $\alpha$ by $\theta$, but $\alpha^\theta$ is that gate raised to the power of $\theta$, where $\alpha = X, Y, Z$.  

It is important to note that the concept of a 'layer' is mainly aesthetic in QVCs, the layer operations are not the same as in classical neural networks. In neural networks a layer indicates that there is matrix multiplication of the inputs and weights, a layer in a QVC merely indicates a group of operations, i.e. after you make the circuit you could change all the layer 'cutoffs' and that would not change the mathematical operations of the circuit. This layer architecture is an expansion upon an design that has been empirically shown to be one of the most powerful QVC architectures \citep{sim2019expressibility}. 

Contemporary encoding approaches in quantum RL fall short of the necessary efficiency for the large inputs Atari requires, as previously indicated. We take a new approach and utilize a neural network to provide an approximate encoding. This idea can be used to enable quantum RL to learn benchmarks that were previously inaccessible due to the large observation spaces. We use a classical neural network to convert the classical pixel data into quantum data. Specifically, this network takes as input the classical data and outputs rotations for the gates in order to establish an approximate encoding. We use the same 3 layers as before to encode the QVC due to their expressibility, i.e. the function space they can learn. Similar to traditional neural networks, for QVCs with large numbers of qubits and many layers, one challenge is the quantum barren plateaus problem \citep{mcclean2018barren}. Barren plateaus is the QML version of vanishing gradients as the number of qubits and the depth of the circuit gets larger. We combat this problem in two ways. First by utilizing QCNN layers which are able to more effectively sidestep barren plateaus in gradients \citep{pesah2020absence}. Second we use an initialization technique, called identity block initialization, specifically designed to combat this problem \citep{grant2019initialization}. This initialization strategy involves selecting the first parameters randomly, then selecting the next parameters to undo the transform. 

\begin{figure}[h]
    \centering
    \scalebox{0.9}{
    \begin{quantikz}[row sep=0.35cm]
    \lstick{$Source$} & \qw & \qw & \qw & \gate{X^{\theta_3}} & \gate{Y^{\theta_4}} & \gate{Z^{\theta_5}} & \ctrl{1} & \qw & \qw & \qw & \qw\\
    \lstick{$Sink$} & \gate{X^{\theta_0}} & \gate{Y^{\theta_1}} & \gate{Z^{\theta_2}} & \qw & \qw & \qw & \targ{} & \gate{Z^{-\theta_2}} & \gate{Y^{-\theta_1}} & \gate{X^{-\theta_0}} & \qw 
    \end{quantikz}
    }
    \caption{Parameterized Quantum Pooling Operation}
    \label{fig:fig2}
\end{figure}
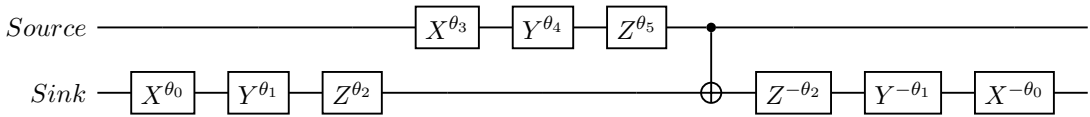

In order to match the action space of the environment we use two techniques: quantum pooling and classical dense. The quantum pooling operation consists of Pauli $X, Y, Z$ gates to a parameterized power, as shown in Figure \ref{fig:fig2}. This operation reduces two qubits into just one qubit; the qubit that is pooled out is called the source and the qubit that remains in operation (and is pooled 'to') is called the sink. We can apply this operation the desired number of times to reduce the number of qubits to the action space, then apply the readout operator and extract the estimated Q value for each action. The classical dense approach involved conducting the same measurements, but then feeding these into a single dense layer, the output of which is the Q-values. We investigate and compare both of these options.

As there are many moving parts in this hybrid model, we present a brief overview of what a single forward pass of the model would look like. A 84x84x4 input is fed into the encoding neural network, as this is the size of framestacked Atari observations. That neural network outputs rotation parameters for 3 layers of the circuit shown in Figure \ref{fig:cir1}, the same structure as the QVC that is applied after the QCNN. These gates are then applied to ground state qubits. The goal of this process is to produce a quantum circuit that is encoded with the pixel information. A quantum convolution is then applied to these qubits (i.e. the unitary circuit shown in Figure \ref{fig:conv} is applied to all neighboring qubit pairs). This serves the same purpose as the traditional convolution operation, specifically for image and spatial analysis. After the QCNN, there are 3 layers of the quantum circuit show in Figure \ref{fig:cir1} applied sequentially to the qubits. Finally, either measurements are made for the qubits and fed into a classical dense neural network, or quantum pooling operations are applied until the number of active qubits is equal to the action space. 

The classical architecture we compare the hybrid to is well established and we use a similar architecture as in \citet{mnih2015human}. Specifically our architecture has 4 convolution operations with 32, 64, 128, 128 filters followed by a 1024 unit and 512 unit dense layer followed by an output layer the size of the action space. Note that this network has on the order of $10^6$ trainable parameters, about 100 times larger than the hybrid model that has $10^4$ parameters. 

\subsection{Experimental Protocol}
The purpose of this study is to empirically evaluate the performance of classical and hybrid quantum-classical DDQN on 2 Atari environments: Pong and Breakout. We chose these environments as they represent two distinct types of game play, with Pong being a multiplayer game in which the agent must learn to play against the default player and Breakout is a single player game, in which the only challenge is the environment (not other players). Quantum techniques have the potential to use less parameters and achieve better policies at a faster rate, as previously stated \citep{lockwood2020, chen2020variational}. We hypothesize that using a neural network to encode classical information into quantum circuits will enable the successful use of QVC's leading to better and faster rewards on these Atari environments

To this end, we propose a total of 130 experiments, 120 hybrid experiments and 10 classical for comparison. For both environments, we repeat each experiment 5 times for both hybrid and classical networks. This results in 10 experiments using the classical neural network as described in Section~\ref{method}. For the hybrid model, there are 12 different configurations, resulting in 12*10 = 120 experiments. See Table \ref{tab:exp} for an outline of all the experiments. 
\begin{table}[h]
\centering
\begin{tabular}{|l|c|c|c|}
\hline
Hybrid Variations & Encoder & Qubits in QVC & Output\\
\hline
D5D & Dense & 5 & Dense \\
\hline
D5Q & Dense & 5 & Quantum \\
\hline
D10D & Dense & 10 & Dense \\
\hline
D10Q & Dense & 10 & Quantum \\
\hline
D15D & Dense & 15 & Dense \\
\hline
D15Q & Dense & 15 & Quantum \\
\hline
C5D & Convolution & 5 & Dense \\
\hline
C5Q & Convolution & 5 & Quantum \\
\hline
C10D & Convolution & 10 & Dense \\
\hline
C10Q & Convolution & 10 & Quantum \\
\hline
C15D & Convolution & 15 & Dense \\
\hline
C15Q & Convolution & 15 & Quantum \\
\hline
\end{tabular} \caption{The 12 Hybrid Variations}\label{tab:exp}
\end{table}  

There are three different aspects of the hybrid model that we experiment with. The first is the encoding scheme. There are two different approaches, classical densely connected neural network layers or classical convolutional layers. Each of these networks will have on the order of $10^4$ trainable variables. This results in a network with two orders of magnitude fewer trainable variables than our traditional approach (as the QVC has on the order $10^2$ parameters). The input into the classical dense layers will necessarily be flattened. The classical convolutional layers are not intended to do the pixel analysis for the QVC. Because in an entangled system, single rotations can change the overall wavefunction, nearby inputs and spatial relations are important considerations for encoding. This is not a simple dimensionality reduction strategy. Our goal is not simply to reduce the information to fewer numbers, but rather to have the neural network learn to convert the information into rotations that can represent the information. Specifically, the input to the parameterized circuit will be the circuit shown in Figure \ref{fig:cir1} with the rotations being the outputs of the neural network, i.e. if there are 5 qubits and 3 layers, then each layer has 24 parameters and there are $3 * 24 = 72$ numbers the neural encoder outputs each one corresponding to one rotation gate. To help elucidate the goal of the neural network encoder, consider a single qubit with a single rotation gate. The one parameter of this gate, $\theta$, can create a state which represents two numbers, e.g. by rotating $\pi/5$ it creates the superposition $\Psi = cos(\frac{\pi}{10})|0\rangle + sin(\frac{\pi}{10})|1\rangle$. The second aspect we vary is the number of qubits, specifically we evaluate using 5, 10 and 15 qubits. Simulation sizes are constrained by the exponentially increasing computational cost. The reasoning behind these qubit choices is straightforward: not all the information present in the pixels is relevant for making informed actions, thus the amount of encoded information may not have to be the full 84 by 84 by 4 array. The computational expense of simulating quantum circuits also exponentially increases with the number of qubits. 5, 10 and 15 qubits have representational power of $2^5, 2^{10}, 2^{15}$ or 32, 1024, and 32,768 respectively. Thus, the 15 qubits are capable of representing the 84*84*4 = 28,224 numbers from the pixels. The third and final aspect is the output of the model. The output can be directly evaluated from the quantum readout operators (after pooling) or a classical dense layer can be attached at the end of the model. We experimented with this idea in \citep{lockwood2020}, and found that the quantum outputs generally performed better. In this work, we expand experiments to QVCs with more qubits and more parameters. 

Finally, the hyperparameters will be held constant across experiments in order to ensure an accurate comparison. Our current set of hyperparameters are outlined here and although they are subject to small changes and optimizations (as is important in machine learning), whatever is done to optimize hyperparameters will be shared across models. These initial hyperparameters are inspired by those used in \citep{mnih2015human} and \citep{andrychowicz2020matters}. The replay buffer is size 1,000,000 with a mini-batch size of 32. For $\epsilon$ greedy exploration the initial $\epsilon = 1.0$ with a decay of $\epsilon_{decay} = 0.99$, $\epsilon_{min} = 0.01$ and a reward discount factor of $\gamma = 0.99$. In addition, both hybrid and classical models use the Adam optimizer \citep{kingma2014adam} with the same learning rate schedule, starting at 0.001 decaying linearly to 0.0001 over 10,000,000 frames. 

With all 130 experiments, this work should provide substantial empirical insight into the use of hybrid quantum classical models for complex reinforcement learning tasks. We hypothesize that the convolutional encoders will perform superior to the dense encoders due to their ability to work with spatial relations, and that all qubit numbers will be able to learn but the best performing will be the 15 qubits because of the ability to represent all the input data (with fewer than 15 qubits, some pixel information is inherently lost), and finally that the quantum output will perform better than the dense. If the qubit encoding performs better with fewer qubits, that demonstrates there is substantial unnecessary information in the input as the fewer qubits can only represent a small fraction of the total input. Therefore, we predict C15Q to perform the best.  

\section{Documented Modifications}

\begin{figure}[h]
    \centering
    \includegraphics[scale=0.6]{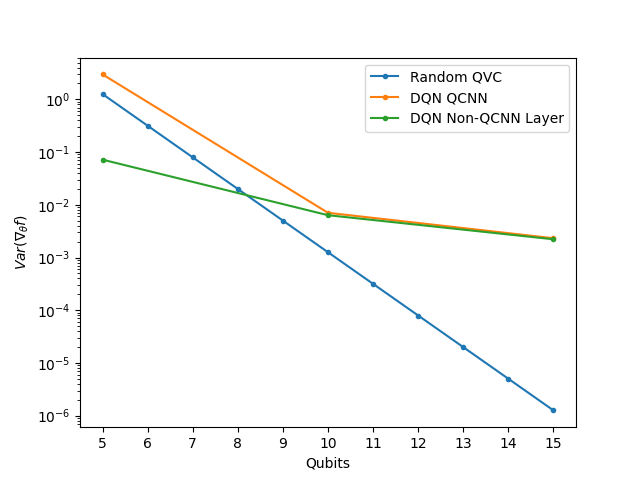}
    \caption{Variance of Gradients in Quantum DDQN and in Random Circuits}
    \label{fig:var}
\end{figure}

Prior to presenting the results, we outline some of the differences from the experiments as presented above and as run. Changes made were small hyperparameter differences and techniques to improve speed and convergence. First, we changed the classical architecture convolutions filters to match the standard Nature DQN \citep{mnih2015human}, i.e. 32, 32, 64 filters. This does not change the total number of parameters (maintaining $O(10^6)$ trainabale parameters). We also slightly modify the structure of the circuit, using two QCNN/pooling layers (instead of one) and and 4 layers of the Figure \ref{fig:fig2} circuit (instead of 6). This maintains the same number of quantum parameters, $O(10^2)$. We did not utilize the initialization strategy as we found it to be unnecessary. Specifically, we found that the variance of the quantum gradients was large enough that the theoretical concern of vanishing gradients was not a problem. See Figure \ref{fig:var} for a comparison of the variance of the gradients in our circuit architectures and the barren plateau variance, with the slope from \cite{mcclean2018barren}. 

Finally, to speedup (and make the experiments feasible) we compute quantum gradients using the adjoint method \citep{luo2020yao, plessix2006review}, which is a technique to accelerate quantum gradient computations. We discussed the Parameter Shift rule to explain how hardware gradients are calculated; however, simulating quantum circuits is substantially more time consuming than running on quantum hardware. We use the adjoint method on simulators to accelerate the gradient calculations by doing operations that are not feasible on quantum hardware. This is merely a technique to accelerate computation and changes nothing about the results (as adjoint and parameter shift methods compute the same gradients) and nothing about the feasibility on real quantum hardware (because real quantum hardware is much faster). 

\section{Results}
\begin{figure}[h]
    \centering
    \includegraphics[scale=0.6]{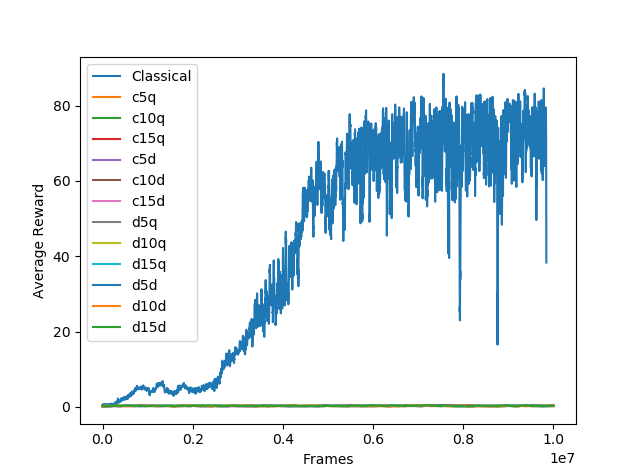}
    \caption{Comparison of Quantum and Classical Methods on the Breakout Environment}
    \label{fig:break}
\end{figure}

\begin{figure}[h]
    \centering
    \includegraphics[scale=0.4]{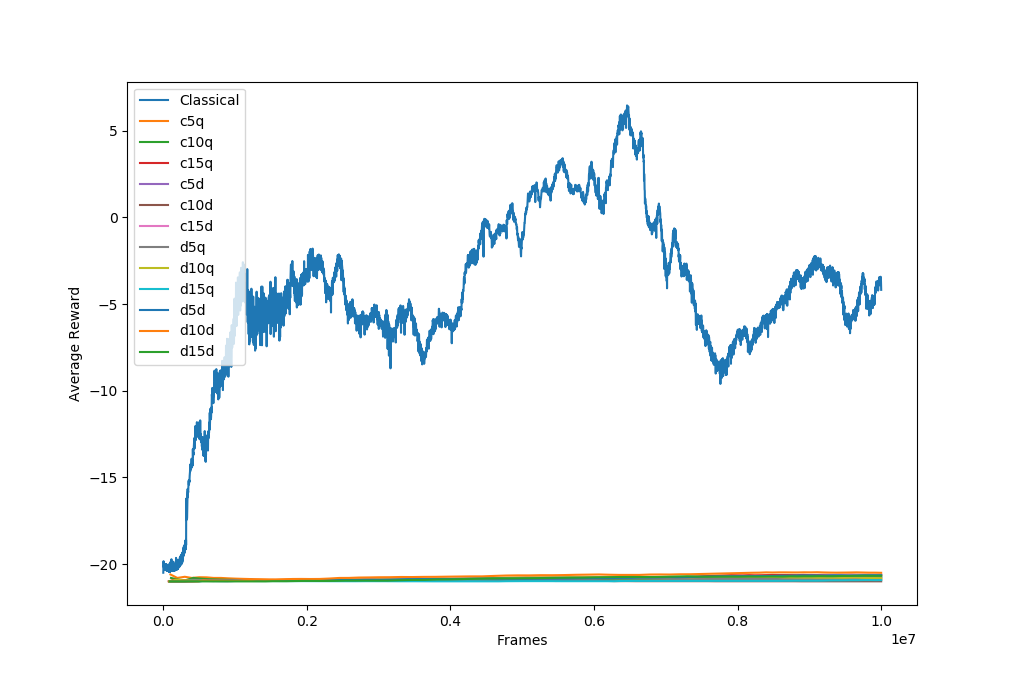}
    \caption{Comparison of Quantum and Classical Methods on the Pong Environment}
    \label{fig:pong}
\end{figure}

With 8 Intel Xeon E5 v3 CPUs full training of a 5 qubit system took about 18 hours, a 10 qubit took about 40 hours and a 15 qubit system took about 65 hours. TensorFlow-Quantum does not use GPUs, hence why we trained exclusively on CPUs. The results are shown in Figures \ref{fig:break} and \ref{fig:pong} as plots of average rewards over the training period of 10 million frames. The results indicate that all 12 quantum variations consistently failed to learn, independent of structure, environment, or random seed. Figure \ref{fig:break} shows the results for Breakout and Figure \ref{fig:pong} shows the results for Pong. While reinforcement learning is notoriously brittle \citep{henderson2018deep, engstrom2019implementation}, we found that Breakout was especially brittle, a finding that aligns with existing implementations\footnote{\href{https://github.com/dennybritz/reinforcement-learning/issues/30}{https://github.com/dennybritz/reinforcement-learning}}. In these figures not every quantum plot is entirely visible, this is because their rewards are so similar the overlap does not show on the plots.

\begin{figure}[h]
    \centering
    \includegraphics[scale=0.5]{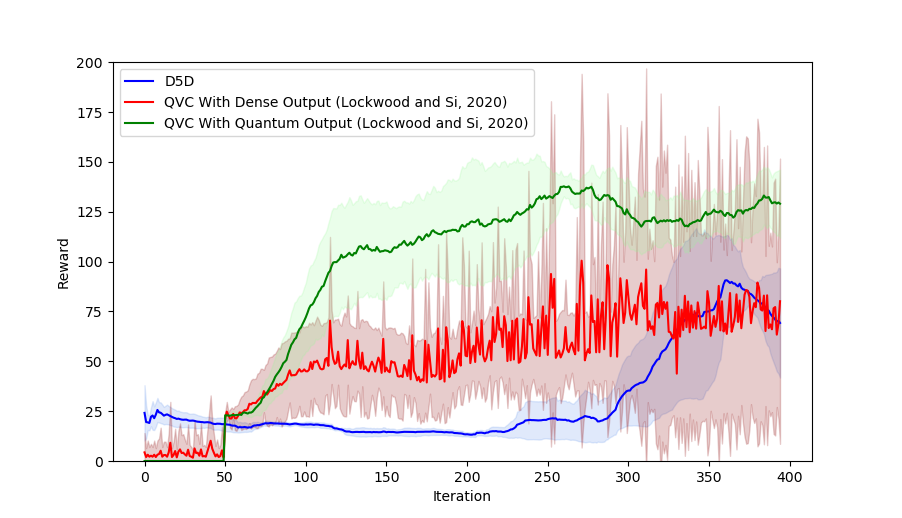}
   \caption{Hybrid-Quantum Classical System D5D on the CartPole Environment}
    \label{fig:cartpol}
\end{figure}

In additional to the proposed experiments we also conducted additional evaluations and comparisons in an attempt to determine the source the hybrid model's performance. To see if our hybrid model was capable of learning at all, we ran tests on simple environments. The results for CartPole are presented Figure \ref{fig:cartpol} and compared with previous approaches from \cite{lockwood2020}. These results indicate that our model is able to learn simple environments similar to previous works. We also experiments with a classical model of comparable size to the hybrid model, specifically with $O(10^4)$ parameters. These results are presented in Figure \ref{fig:small}. These results establish the common knowledge idea that very small classical models are unable to learn anything in the Atari environments. 

\begin{figure}[htbp]
\floatconts
  {fig:small}
  {\caption{$O(10^4)$ Parameter Neural Networks on Atari Environments}}
  {%
    \subfigure[Pong][c]{\label{fig:small_pong}%
      \includegraphics[width=0.45\linewidth]{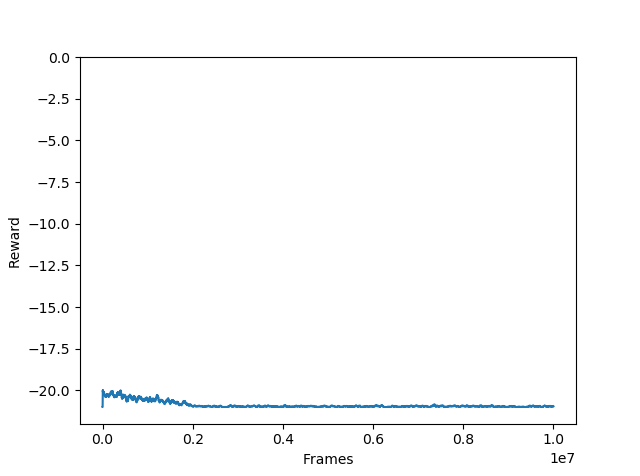}}%
    \qquad
    \subfigure[Breakout][c]{\label{fig:small_bb}%
      \includegraphics[width=0.45\linewidth]{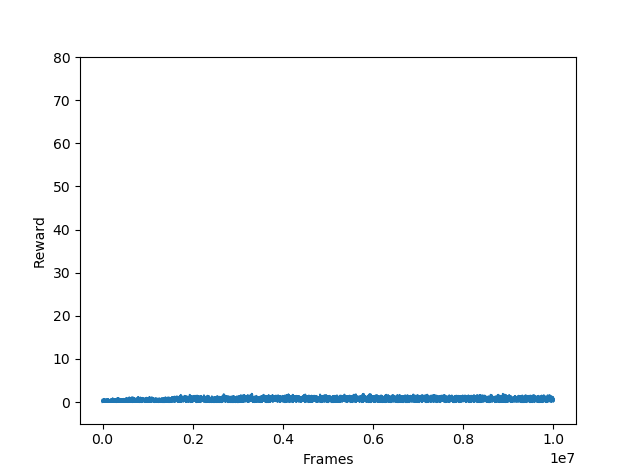}}
  }
\end{figure}

\section{Findings and Discussion}

We argue that these results stem from a lack of function approximator expressibility. It may be tempting to ascribe the failures of learning to be fundamental within the system. I.e. that the setup we constructed is flawed in a fundamental way such that it cannot learn Atari because it cannot learn anything. However, we know this is not the case as on much simpler tasks the hybrid model demonstrates the ability to learn as shown previously in Figure \ref{fig:cartpol}. Given the success on small scale environments, this indicates that the hybrid model isn't fundamentally flawed but is insufficient for the complex Atari environments. Just as a $O(10^4)$ MLP fails to learn anything on the Atari benchmark, as previously shown in Figure \ref{fig:small}, the hybrid model is likely too small (since the hybrid model is about 100 times smaller than the classical model). Our ansatz (i.e. the circuit structure) may also be limiting the model, as the cost of reducing the problems of barren plateaus is reduced expressibility \citep{holmes2021connecting}. 

While recent results have expanded to show more successful quantum RL agents \citep{skolik2021quantum, jerbi2021variational} they remain limited to simple environments (e.g. CartPole). However, \cite{skolik2021quantum} indicated that data re-uploading \citep{perez2020data} was important to successful value based quantum RL. Data re-uploading involves repeating the encoding step prior to every layer. This continued trend of consistently positive and advantageous results inspired this work to investigate environments much more complex than previously explored. However, this increase in complexity may require much more advanced techniques than the quantum RL community currently possesses. 

These results generally fall in line with recent findings in quantum machine learning. Specifically, results that counter the idea that quantum machine learning is a panacea and that QML likely has advantages only on specific problems in specific situations. \cite{huang2020power} showed that it is unlikely that QML will offer advantages unless the data has something sufficiently \say{quantum} about it, such as being generated from quantum circuits. \cite{huang2021information} proved that for quantum processes, QML provides no advantage in minimizing the average prediction error, only providing an advantage for minimizing the worst-case prediction error. \cite{kubler2021inductive} and \cite{qian2021dilemma} showed there is little indication that QML can improve supervise learning. 

\subsection{Future Work}

This negative results present here strongly encourage future work in this field. There are a number of techniques that could be investigated that may result in better results. Our models do not utilize data re-uploading, which may influence their results and applying this technique may unlock performance gains. Other ideas include trying different encoding strategies, for example amplitude encoding. This is not a feasible for near term hardware, but is still worth investigating whether it can yield positive results. There is also room for more hyperparameter optimization. Given the differences between classical and quantum learning, in addition to the brittleness of RL, it may be that substantially different hyperparameters are necessary for quantum RL to succeed. This negative result does not indicate a dead end, but the opportunity for many new techniques. 

\section{Conclusion}

In this work, we expanded upon previous works in quantum RL and designed large scale experiments of hybrid quantum-classical reinforcement learning. We evaluated these agents on a subset of the Atari benchmark, using neural network encoders to enable observation spaces thousands of times larger to be processed. We evaluated 12 different hybrid agents and found that they consistently failed to learn, underperforming classical methods. Through the results presented here, we argue that further advancements are necessary in the field of hybrid quantum-classical reinforcement learning is to master contemporary benchmarks.  

\acks{The authors would like to thank Michael Broughton for their advice and assistance.}

\newpage

\bibliography{refs}

\end{document}